# E2XLRADR (Energy Efficient Cross Layer Routing Algorithm with Dynamic Retransmission for Wireless Sensor Networks)

Kanojia Sindhuben Babulal, Rajiv Ranjan Tewari

Department of Electronics and Communication
J.K Institute of Applied Physics and Technology
University of Allahabad, 211002
Email: sindhukanojia@gmail.com

### *ABSTRACT*

*The main focus of this article is to achieve prolonged network lifetime with overall energy efficiency in wireless sensor networks through controlled utilization of limited energy. Major percentage of energy in wireless sensor network is consumed during routing from source to destination, retransmission of data on packet loss. For improvement, cross layered algorithm is proposed for routing and retransmission scheme. Simulation and results shows that this approach can save the overall energy consumption.*

### *KEYWORDS*

*Cross layer optimization, Wireless Sensor networks, and protocols*

## 1. INTRODUCTION

A wireless sensor network (WSNs) is an infrastructure comprising of sensing, computing, and communication elements which gives as administrator the ability to instrument, observe, and react to events and phenomena in a hostile environment [1]. Wireless sensor network comprising of tiny power constrained nodes and limited battery support are gaining popularity because of their peculiar advantages and vast range of opportunities they enable. Four major components of a sensor network are: (1) assembly of distributed sensor; (2) interconnecting networks; (3) a central point of information clustering; and (4) set of computing resources at the central point [2].

Cross layer design is currently one of the most active research areas in computer networks. Cross layer interaction means allowing communication of layers with any other possibly non-adjacent layers in the protocol stack [3]. Traditionally, network protocols are divided into several independent layers. Each layer is designed separately and the interaction between layers is performed through a well defined interface. The main advantage of this is architectural flexibility. Cross layering was thought of to address QoS(Quality of Service), energy consumption, poor performance, wireless links, mobility, packet loss, delay problems observed in wireless networks. It tries to share information amongst different layers, which can be used as input for algorithms for decision processes for combinations and adaptation. This process of sharing has to be co-ordinated and structured somehow since cross layering could potentially worsen the performance problem that it intends to solve. But it should be kept in mind that cross layering is not simple replacement of layered architecture nor is it the simple combination of layered functionality.

The power sources of sensor nodes have limited battery budgets. Lots of energy is consumed in data transfer and its processing. The major factors responsible for power consumption in wireless sensor networks are collision, overhearing, control packet overhead, idle listening and over emitting. If one can reduce any of the above problems one can reduce power consumption to some extent. The main characteristic features of sensor networks are that sensor networks are densely deployed and are prone to failure. The topology of sensor networks changes very frequently and they have limited power, computational capacities and memory. There are number of challenges to overcome for a wireless sensor network to become truly ubiquitous. These challenges and hurdles of wireless sensor networks include, but not limited to the following: limited functional capabilities, power factors, node costs, environmental factors, transmission channel factors, topology management complexity and also scalability problems. Of all these factors, the paper addresses the issue of the power factor.

## 2. RELATED WORK

In most recent literature, improving the lifetime of the network, throughput and delay have been investigated as a key measure of the network performance. Most of the research work has been focused on improvement of the lifetime of the network. There are many algorithms and protocols developed[4,5,6,7,8] which try to reduce energy consumption by periodic listening and sleeping, collision and overhearing avoidance and message passing etc. But there are still many things to be done for improving these protocols and algorithms because they were designed independently based upon conventional layer network which have several drawbacks in terms of performance and efficiency of the system. For wireless sensor a traditional layer network design is not very efficient. In this paper we address the issue of cross layer design applying the knowledge of wireless medium of physical layer and MAC sub layer being passed to the network layer and the information of network being transmitted to the lower layers. Many papers can be found on the trade off between throughput and delay. Paper [9,10] talks about high throughput and low delay in wireless networks. Paper [11] gives details about the delay throughput only for mobile ad- hoc networks. In [12] analysis of delay-throughput for various wireless network topologies has been done. In [13] author has proposed a new geographic routing scheme by which a near optimal capacity is achieved with low delay. Authors in [14] have developed an algorithm that shows delays in the packet delivery. In Paper [15] a cross layer algorithm which takes the advantages of the retry limit has been given. In [16], for the delivery of the packets, a retransmission deadline is assigned to each and every packet by the application layer. A.Sivagami et al in [17] have used collection tree protocol (CTP) to collect data from the sensor node. It uses either the four bit link estimation or Link estimation Exchange protocol (LEEP) to predict the bi-directional quality of the wireless link between the nodes and the next hop candidate is based on the estimated link quality. The residual energy of the node is an important key factor, which plays a vital role in calculating the lifetime of the network. S.Mehta et al in paper [18] have proposed a scheme to save power dividing the execution of back off algorithm by avoiding idle listening. In paper [19] algorithm makes the routing decision using the statistics of the energy consumed for each type of node activities including sensing, data processing, data transmission as a source node and routing process.

The rest of the paper is organized as follows. In section 3 model of wireless sensor network is given; in section 4 we propose an energy efficient routing scheme; in section 5 a new retransmission scheme is discussed; section 6 shows simulation results and finally section 7 concludes the paper and discusses the future work.

## 3 – NETWORK MODEL

We model the dynamic wireless model as a set of N nodes distributed into two dimensional planes. Sensor nodes are deployed with high density, the sensing range of each node is overlapped with some other nodes. All the sensor nodes should be fixed or should have less mobility. As wireless sensor networks do not have fixed infrastructure, sensor nodes collaborate to work together by wireless channel. In multi-hop sensor networks, each node plays the dual role of data originator and data router when two nodes cannot directly link each other.

In this paper we make the following assumptions.

- Nodes use the same frequency for transmitting with an Omni-directional antenna (the node has one receiver and one transmitter which cannot transmit and receive at the same time).

- Each node has information about other nodes in its sensing range.

- All the nodes are fully charged initially.

- Two nodes i and j can transmit a packet to each other with a transmission power pi with the range $0<i<Pmax$, where Pmax is the maximum transmit power of the node.

## 4 – E2XLRADR (Cross layer Routing Algorithm with Dynamic Retransmission Scheme)

We use a cross layer optimization between PHY, MAC and Network layer for routing. For a given path between source and the destination each intermediate node computes a new limited number of retransmission. This parameter can be adjusted easily by each node. Network layer maintains routing algorithm. So each node acts as router, it permits to relay packets originated from source S to destination D. It must carry routing information which permits sending of packets to destination via a neighbor. Using the concept of sending the data to farthest distance intermediate node is same as in [20].

A packet fails only when there is interference on the intended receiver. The only source of packet loss is due to collision. Even if the packets collide partially they are considered to be collided. For a reliable communication, we allow a limited number of successive transmissions for a packet, after that it is dropped. We assume that throughout the process there is some mechanism that notifies the sender of success or failure of its transmission. XLRA is divided into 5 major phases as stated below:

### 4.1 Finding the Destination Location.

In the first phase when the source S has data to send to Destination D, Source S checks where Destination D is, whether it is in its sensing area or not. If D is in sensing range then S sends an r-request msg to know whether D is in position of receiving the data or not. If D is in the state of receiving the data i.e. it is neither receiving not transmitting, D sends back the acknowledgement ack1 msg that "I am ready to receive the data you can send". Then S sends data directly to D; after successfully receiving data D sends ack2 msg to confirm that data is received. If D is not in the receiving state, S waits for D to be in the ready state. If S does not receive the acknowledgement ack1 msg within the fixed time Tf it resends the r-request msg. Tf is fixed .As stated before while a communication is going on other nodes can transmit or receive. Other nodes in the sensing range can sleep for this communication time period. Discussion about the sleeping criteria in detail is out of scope of this paper.

Algorithm 1

1. Source S has data to send to destination D.
2. S checks for the position of D, whether it is in its sensing range or not.
3. If D is in sensing range, S sends r-request message msg to D saying that it has data to send.
4. If D is ready to receive data it sends back acknowledgement ack1 msg to source S saying that it is ready to receive data.
5. Source sends data to destination D.
6. Destination D sends ack2 msg to confirm that data is received
7. If D is not ready to receive data, source S waits for Tf amount of time
8. And again sends r-request to D
9. If S does not receive ack1 in time Tf, S retransmits r-request msg

**4.2 Route Finding**

In the second phase when a node has data to send to and the destination D is not in its sensing range and no prior knowledge of routing to destination is available in its cache, source node issues an r-request to the node at the maximum distance in its sensing range. The length of r-request is small to minimize the energy required. Then that maximum distance intermediate node Mdi issues an acknowledgement ack1 msg confirming that it is in the ready state to receive the data. Then Source sends the data to that intermediate node. If the ack1 msg is not send within the time Tf, the source retransmits the r-request msg. Suppose there are two or more than two nodes at the same maximum distance then S sends r-request to all of them. Now all of them will reply with ack1 msg and with the information about the power level of it. When source receives msg it sends the data to that node that has maximum energy. When the data is send to intermediate node it has to send back another ack2 msg that it got the data, which confirms that the data is received. Other node can go to sleep mode to save energy. Now it's the job of intermediate node to send the data to its maximum sensing range. And in this way the data is send to the destination node. In this way the risk of data loss is less as in all the phase acknowledgement msg is send back. This can be done using RTS/CTS protocol.

Algorithm 2

1: Source S has data to send to destination D

2: Checks algorithm 1

3: If no prior information about routing available in cache

4: S issues an r-request msg to the maximum distance intermediate node Mdi where i=1,2,3……n

5: If Mdi is in ready state it sends back the ack1

6: if ack1 is not received in time Tf then it retransmits r-request

7: if ack1 received with in Tf, S sends the data to Mdi

8: Mdi sends ack2 confirming that data has been received.

9: If more than one node is at the same maximum intermediate distance, then S issues r-request msg to all of them Mdn1, Mdn2, Mdn3…...Mdnn

10: All those nodes reply with ack1 msg and information about their power level.

11: S checks the maximum power level node and sends data to that particular node.

12: Mdn replies with ack2 msg for confirmation of data received.

13: If within Tf, ack1 is not received, r-request is retransmitted.

14: Intermediate Mdn repeats step1 to step 12.

**4.3 Dynamic Scheme for retransmission for maximum number of transmission**

In this section we propose a new dynamic Kmax(maximum number of retransmission) limited algorithm based on a table driven routing where routes are already known. It is a cross layer scheme where each node needs the information about the route and the number of hops to determine the value of Kmax. These two pieces of information can be provided from the routing protocol. If they are available, our scheme computes the corresponding Kmax. We use a simple method to calculate the dynamic value of Kmax. Let us take a small example to understand this scheme.

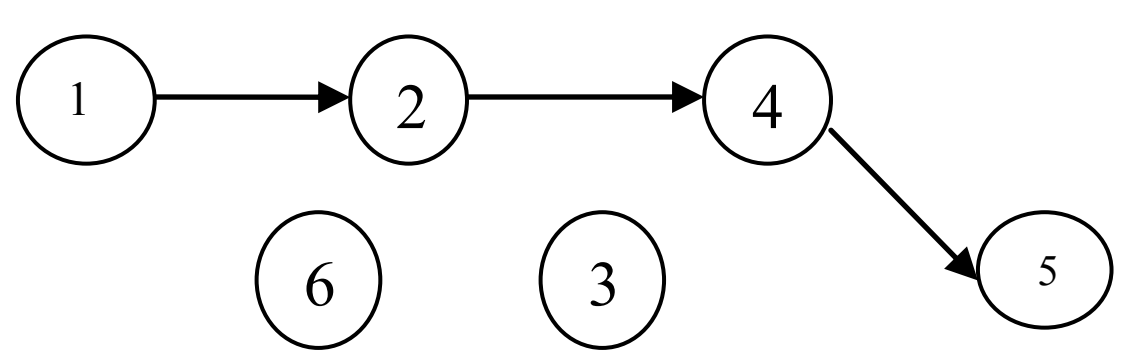

Fig 1. Example of data transfer

In above cited fig 1, let us suppose Source S (1) has data to send to Destination D (5). Suppose the data has collided at node 2. Now node 1 has to retransmit the data to 2 again, then 1 checks the number of hop (m) to the destination which is 3. So initially Kmax is set to 3.Three times the packet would be retransmitted otherwise the packet would be dropped. At the next intermediate node this value would be increased to (m+1), then to (m+2) and so on. An important thing to note is that at the starting of the process for retransmission Kmax is set depending on the number of hop. Suppose the failure is at the any intermediate node say node 4, it will check the initial Kmax and calculate the new Kmax' depending on its position from source and the number of hops left from the destination. To find Kmax' the packet failure node will check its position from source and hops left from the destination.

Kmax' = (position of source) + (number of hops left from destination)

In case of node 4

Kmax' = (2 + 1) = 3

Under this scheme, we aim to give more chance of success to packets that had come near to their destination. It rather means that we need to avoid as much as possible losing packets near their destination, so that waste of bandwidth throughput on a path becomes lower. Normally, the way of choosing a good Kmax should depend not only on the number of hops, but also on factors like transmission probabilities and number of neighbour. But taking care of many parameters at the same time is a complex issue.

Algorithm 3

1. If S does not receive the msg ack2 within Tf time
2. S has to retransmit the data
3. S checks the number of hops to the destination node(H) say m
4. Depending on the number of hops, Kmax( maximum limit of retransmission) is set to m ie Kmax=m
5. m increased to (m+1) at the next intermediate node, and (m+2),(m+3)…and so on depending on the number of hops
6. if failure is at other intermediate node, that node calculates new Kmax' from the intial Kmax
7. Kmax' = (position of source node) + ( number of hops left from destination)

### 4.4 Route Maintenance

In wireless sensor networks, sensor nodes are often mobile and topology of network changes frequently. But it is assumed that this algorithm is best suited to less mobile network. Each maximum distance intermediate node in route is capable enough to monitor its potential increase of interference and decrease in residual energy level. When node cannot find its next maximum distance node it sends route recover message.

### 4.5 Route Re-establishment

Route re-establishment is necessary when condition of sensor networks change greatly and route maintenance is not able to recover the lost link. This decision is only dependent on the potential interference and residual energy of the current node.

## 5. SIMULATION AND RESULTS

The cross-layer algorithm described above was implemented and evaluated in OPNET simulator [21]. Here we describe about the behaviour of our algorithm. We used the discrete event driven simulator Opnet for our simulations. The set up consists of simulation of 50 sensor nodes which are randomly deployed in a rectangle area 1000 * 1000 $m^2$. Five of them are chosen as the source node, which produce sensing data. Other 45 nodes are data sink node. We use network lifetime as metric to evaluate the performance of our cross layer design. We define network lifetime as time taken for 30% of sensor nodes in the network to drain up their power. We change the mobile rate to show the performance of E2XLRADR. Figure 1 to Figure 3 shows the network lifetime of our approach and reference DSR. As the mobile rate increases, the lifetime of DSR changes, while E2XLRADR can alleviate this situation. It shows that E2XLRADR significantly increases the lifetime of network. A network lifetime increased as much as 30 % of the lifetime of a DSR networks in low mobility scenarios. Figure 4 shows that when the delay is less, the through-put increases significantly because retransmission gets decreased. Simulation parameters are listed in table 1.

| | |
|---|---|
| Transmission Range | 250m-350m |
| Network Size | 1000 * 1000 $m^2$ |
| Node Number | 50 |
| Speed of nodes | 3,6,9 packets/second |
| Simulation Time | 70000 |
| Packet Size | 1024 bits |

Table 1: Simulation parameters

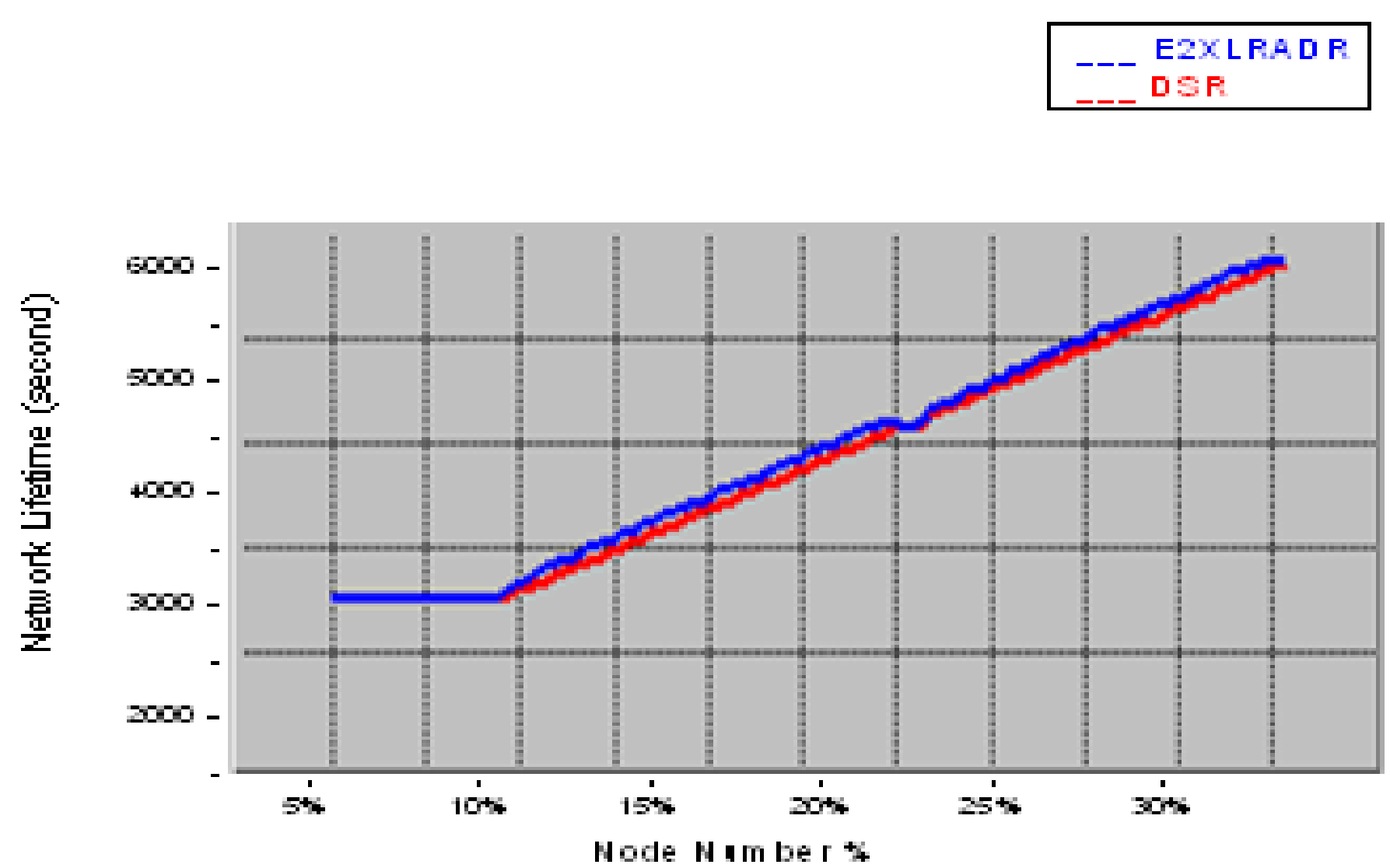

Figure 1. Network Lifetime of two different schemes at mobile rate 3 packets/second

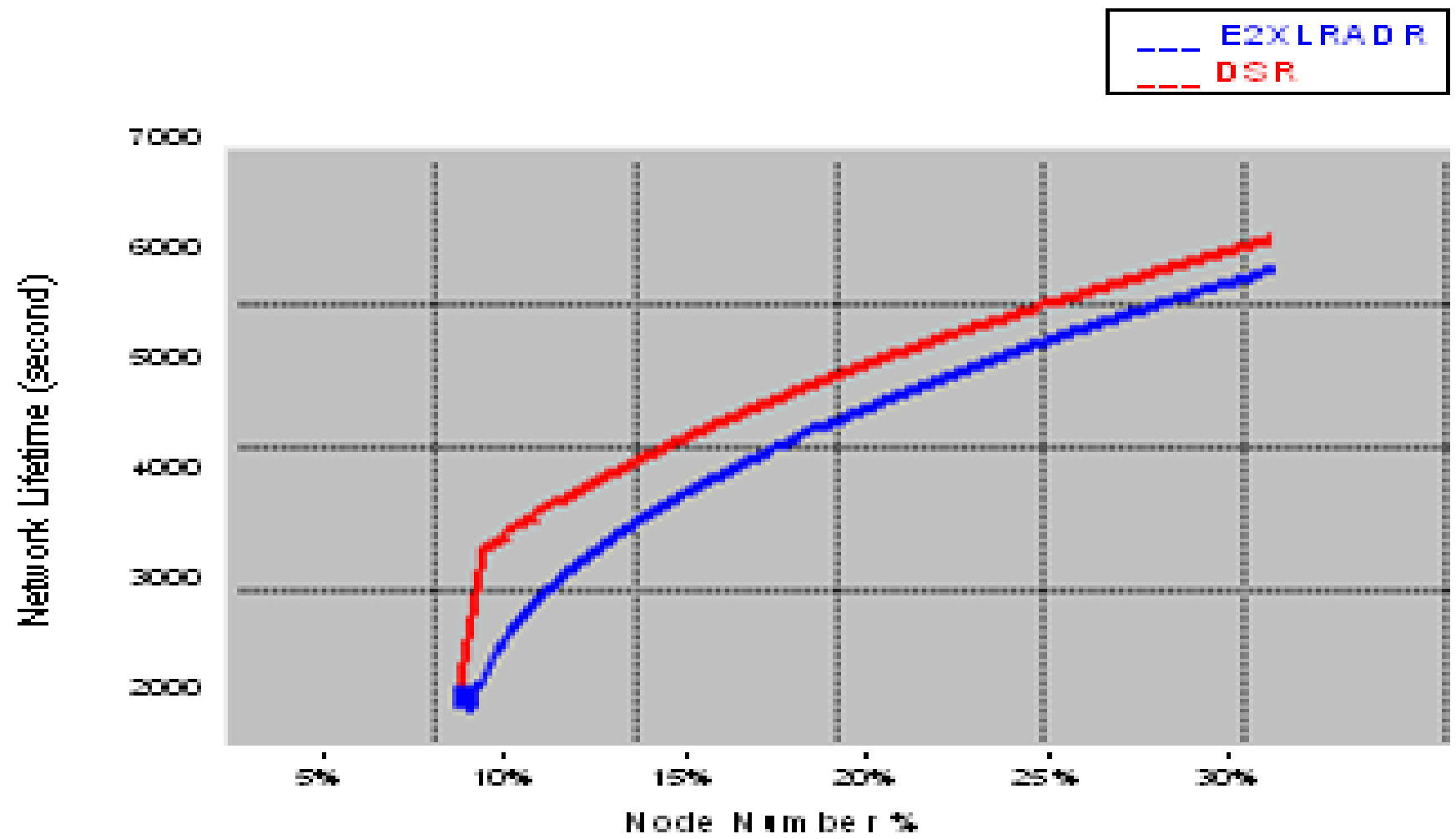

Figure 2. Network Lifetime of two different schemes at mobile rate 6 packets/second

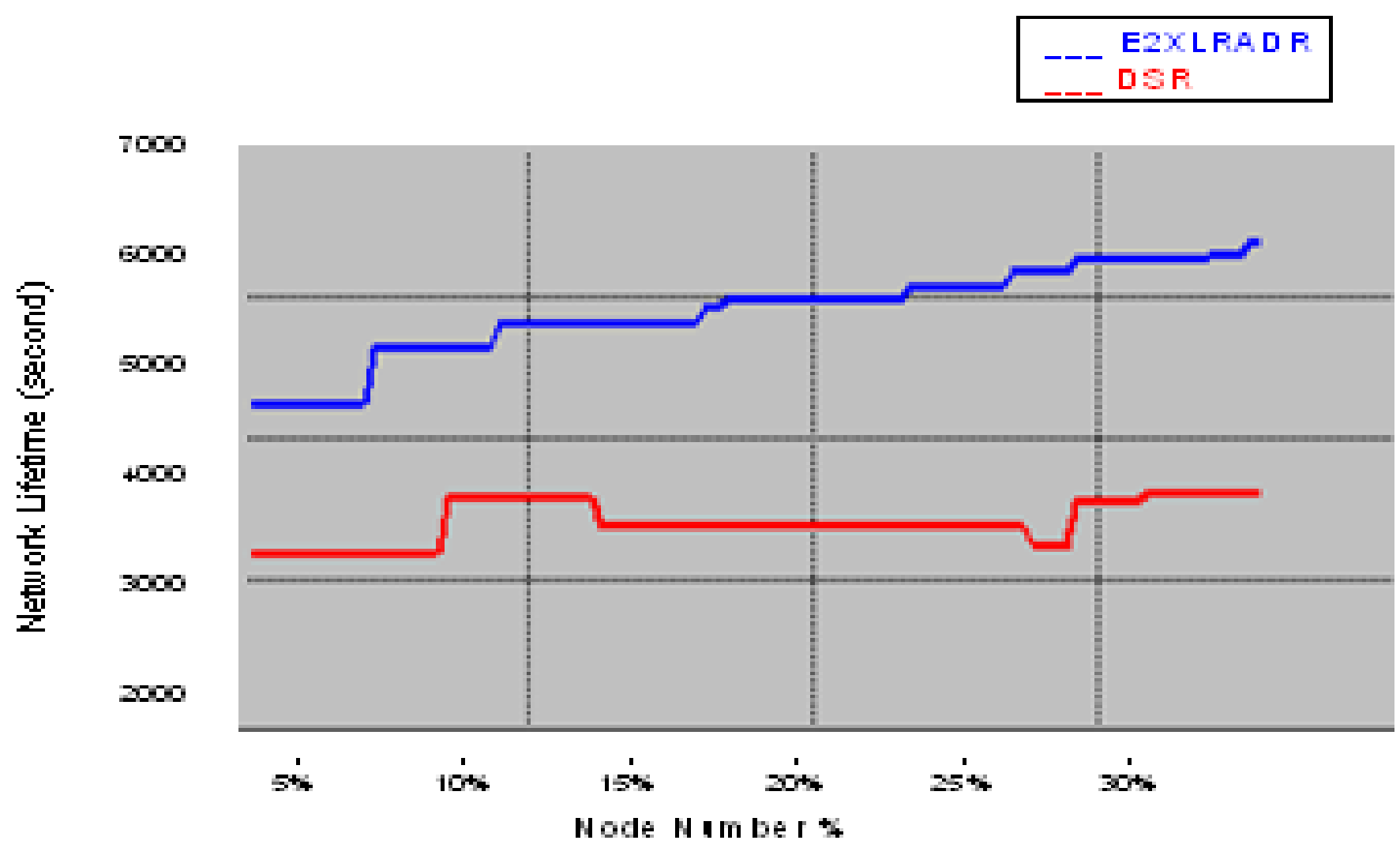

Figure 3. Network Lifetime of two different schemes at mobile rate 12 packets/second

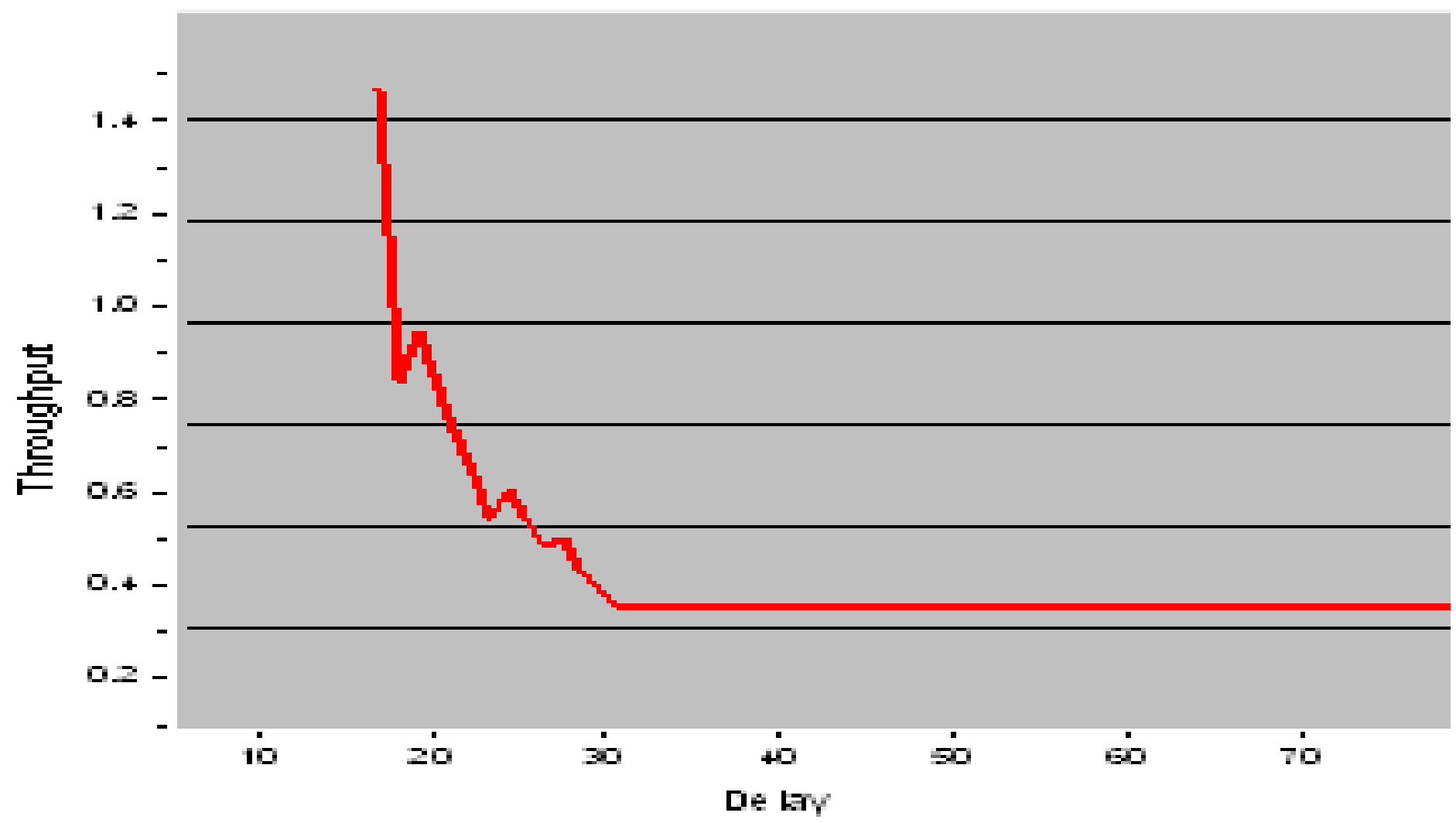

Figure 4. Throughput is high when delay is less

## 6- CONCLUSION AND FUTURE WORK

Cross Layering is the best approach to save energy in wireless sensor networks. Energy efficiency can be improved at various layers. The knowledge of physical MAC, and network layer should be shared with each other properly. The conventional layered approach has several drawbacks in the system design. In this proposed scheme, we have used three main concepts, sleep/awake, limited number of dynamic retransmission, and farthest intermediate node. In summary it may be concluded that dynamic scheme provides better performance when some of the source nodes collaborate by forwarding packets and when these kinds of sources are well distributed in the network. The limited number of re-transmission has direct impact on the end-to-end throughput and delay. Hence our scheme saves energy and it also concentrates on the throughput and delay. In future we aim to verify the process of proposed solution experimentally.

## Authors Biography

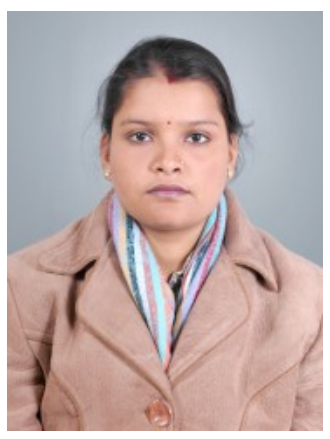

Kanojia Sindhuben Babulal (sindhukanojia@gmail.com) is pursuing her D.Phil. in Department of Electronics and Telecommunication at J.K Institute of Applied Physics and Technology, University of Allahabad at Allahabad. She obtained her M.Sc. (C.S) degree in 2006 from the University of Allahabad, and her B.Sc. in Computer Science from South Gujarat University, Surat, in 2004. Her research is focused on Cross Layer Design in wireless sensor networks, MANETS.

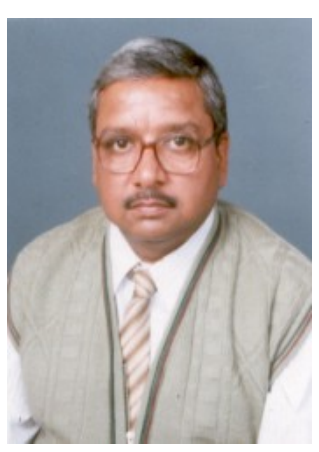

Prof. Rajiv Ranjan Tewari (rrt_au@rediffmail.com), born on May 23rd 1956 at Lucknow (India), is teaching graduate and post-graduate students of Comp. Science/Engineering at the University of Allahabad since 1982. He was awarded K.S. Krishnan Award by IETE in the year 1987 for the best paper on System Design. He is life fellow of IETE. His areas of interest are Real-time System Design, Multimedia, Computer Networks and Robotics.